\documentclass[prl,preprint,aps,showpacs,longbibliography,lengthcheck]{revtex4-1}

\usepackage{graphicx}%
\usepackage{dcolumn}
\newcommand{\jyu}{ Nanoscience Center, Department of Physics,  University of Jyv$\ddot{a}$skyl$\ddot{a}$ ,P.O. Box 35, FI40014 Jyv$\ddot{a}$skyl$\ddot{a}$ , Finland}

\begin{document}

\title{Linear current fluctuations in the power-law region of metallic carbon nanotubes}
 
 \author{D. Talukdar}
 \email{deep.d.talukdar@jyu.fi}
 \altaffiliation{Presently at: 
 Saha Institute of Nuclear Physics, 1/AF-Bidhannagar, Kolkata - 700064 }
 \author{P. Yotprayoonsak}
\author{O. Herranen}
\author{M. Ahlskog}
 
\affiliation{\jyu}

\begin{abstract}

We study low frequency noise in a non-Ohmic region of metallic single walled and multiwalled carbon nanotubes. The generalized relative noise appear to be independent of applied bias in the power-law regime of the tubes and in agreement with theoretical predictions. Beyond the power law regime the suppression of conductance due to scattering with optical phonons is accompanied by a reduction of relative noise by an order of magnitude. Mobility fluctuations in the tubes due to optical phonon scattering cause the unexpected reduction in the relative noise magnitude which is modeled using a modified mobility fluctuation picture. The findings have important implications for metallic nanotubes being used as interconnects in nanoelectronic devices.

 \end{abstract}


\maketitle
Analysis of resistance noise in low dimensional and disordered systems which are easily driven into the non-linear regime is an old but unresolved problem. In this regime, where the resistance becomes a function of applied bias, the analysis of noise goes beyond the realm of fluctuation-dissipation theorem\cite{Kogan} making the analysis through conventional means arduous. The dearth of literature dealing with noise in the non-linear regime can also be partly attributed to it. Theoretical efforts in the past by Rammal \textit{et. al.} \cite{Rammal} first mooted the idea of analyzing noise in the non-linear regime.  Cohn's theorem was used by the authors to derive exponent inequalities in  Charge Density Wave (CDW) systems and metal-insulator composites having $I-V$ characteristics of the form of a power law. However, no experiment has explored the behavior of noise in this regime despite the fact that in power-law  regime noise analysis is simplified compared to other non-linear regimes (e.g., where conductance is expanded in a Taylor series of dc bias voltage). In nanoscale objects due to widespread non-linear $I-V$ properties, especially power-law $I-V$ relations \cite{Zhou,Liu1,Venkataraman,Bockrath}, the measurement and interpretation of noise in this region have become increasingly significant.

         For performing noise experiments in the non-linear regime metallic carbon nanotubes (CNTs) can be used as good test samples as they display  power-law variation of conductance with bias voltage at low temperatures. The power-law dependence in metallic Single Walled nanotubes (SWNTs) is widely believed to be a signature of \textit{luttinger-Liquid} (LL) behavior\cite{Bockrath,Kane} while in metallic Multiwalled nanotubes (MWNTs) the origin has been ascribed to a \textit{environment-quantum-fluctuation} process \cite{Tarkiainen}. In this Letter,  we measure and analyze noise in the power-law regime and beyond in carbon nanotubes. The objective is to show that the generalized relative noise in a power law regime remains independent of applied bias thereby making it possible for fluctuations in this regime to be analyzed in a manner similar to that in Ohmic regime. To illustrate the usefulness of this analysis we show that by measuring noise beyond the power-law regime Optical Phonon (OP) scattering reduces the magnitude of relative noise from the constant value in the power law regime.  We propose a modified mobility fluctuation model which seems to explain the  noise behavior beyond the power law regime in SWNTs. Noise in the power law regime is also measured and analyzed for MWNTs where a larger, cleaner power law region exists with negligible OP scattering effects.
 				 				 
       In the Ohmic regime the resistance noise ($\delta R$) manifests itself in the fluctuations of current/voltage ($\delta I$/$\delta V$) and the current/voltage is merely used to make the fluctuations 'visible' while playing no role in the production of fluctuations \cite{Weissman,Kogan,Talukdarrsi}. The relative noise power ($A_X$) can be determined using a generalized version of Hooge's empirical formula \cite{BardhanUN,Hooge}:
\begin{equation}
A_X= \frac{S_X}{X^2}=\frac{V^{\gamma_o}}{f^\lambda}\Re(R)
\label{eq:hooge}
\end{equation}      
where \textit{X(=R,V,I)} is the fluctuating quantity and  $S_X=\left\langle \delta X^2\right\rangle$ is the spectral density. $R$ is the chordal resistance defined as $V/I$. $\gamma_o$=0 for equilibrium resistance fluctuations and $\gamma_o$$\neq0$ for a driven phenomena. The function  $\Re(R)$ depends upon the system under consideration. In the Ohmic regime, the relative current fluctuation is independent of bias i.e., $\delta I^2/I^2=constant$ as resistance $R$ is independent of the applied voltage.  For the non-linear regime resistance and hence fluctuations \cite{Bardhan} become a function of the applied bias i.e., $\delta R=\delta R(V)$ making the estimation of noise difficult. However, situation is different in case of  $I-V$ relations of the form \cite{Rammal},  \textit{I = g $V^\alpha$} where $\alpha > 0$ and $g$ is some generalized conductance defined as $I/V^\alpha$. In such a case, if $V$ is kept constant during an experiment, then current fluctuations in the sample will arise from fluctuations in $g$, i.e. 
\begin{equation}
\delta I  =  \delta g.  V^\alpha 	  
\end{equation}
The generalized relative noise, $A_I$, can then be written as:
\begin{equation}
 A_I=\frac{\delta I^2}{I^2}= \frac{\delta g^2}{g^2}.
\label{eq:pwrco}
\end{equation}
Thus, in the power law regime the relative noise, determined using suitable normalization and generalized Hooge's relation, is independent of bias. Any bias dependence of $g$ can be taken into account by using the suitable normalization $\delta I^2 / I^{2+\gamma_o}$. Thus we arrive at a non-trivial situation where  noise in the power-law (non-linear) regime can be determined in  a manner similar to that in the linear regime. This will form the foundation through which we analyze our experimental results.

\bigskip
\begin{figure}
\includegraphics[height=6.5cm]{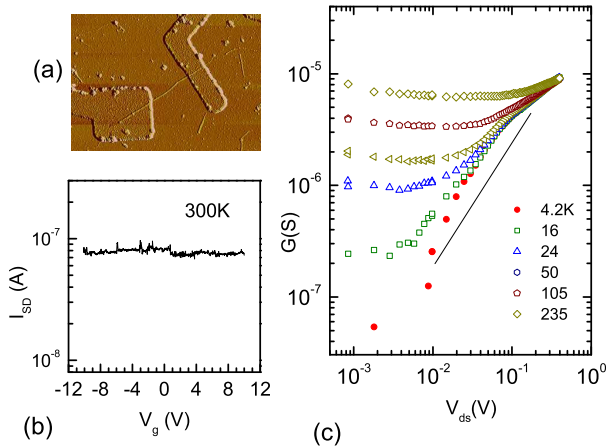}
\caption{(color online) (a) The tapping mode AFM images of a metallic  SWNT of channel length 700 nm and diameter 2.9 nm  with  top contacted Pd electrodes. (b) The source drain current $I_{ds}$ as a function of gate voltage $V_g$. The metallic character of the tube is evident from the gate response. (c) The conductance G versus drain source bias $V_{ds}$ at different temperatures. At 4.2 K the conductance is a true power-law within the measurement range.}
\label{fig.1}
\end{figure}

				The SWCNT and MWCNT devices investigated in this study were fabricated on  top of 300nm  SiO$_2$  layer thermally grown over highly doped Si wafers using standard electron beam lithography. The suspension of commercial SWCNTs used were produced by NanoCyl S.A. (Sambreville, Belgium) and had an average diameter of 2 nm. MWNT samples were obtained from Iijima group.  The suspensions were then randomly deposited onto the chips. Subsequently CNTs were located using atomic force microscope (AFM) followed by metalization using a 25 nm thick Pd layer followed by liftoff. Pd was chosen to improve the contact resistance. The transport measurements were done at several temperatures varying from room temperature down to liquid helium temperature. The low frequency noise measurements were made using a current preamplifier (Ithaco 1211) followed by a low pass filter. The acquired time series was stored and further processed using MATLAB. All measurements were performed inside a RF-shielded room.
			
\begin{figure}[htbp]
\includegraphics[width=9cm]{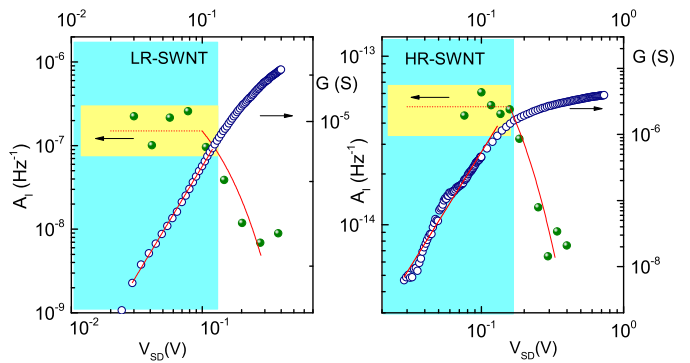}
\caption{ (color online) Relative noise ($A_I$) and conductance ($G$) versus $V_{ds}$  for (a) LR-SWNT and (b) HR-SWNT at 4.2 K. The relative noise is a constant in the power-law region for both the tubes (shown by the yellow band) and is represented by the dotted line as a guide to the eye. The fit by bold line to the noise  using the expression given in the text.}
\label{fig.2}
\end{figure}			 
			 
     		To characterize the CNTs, we first obtain the gate response  as well as the $I-V$ characteristics of the device. The AFM image of a typical CNT device is shown in Fig 1(a). The metallic character of the nanotube is evident from the measured gate response curve in Fig. 1(b). In Fig. 1(c) the typical conductance ($G$) vs drain-source voltage ($V_{ds}$) curves for a metallic SWNT at different temperatures are shown. One can see that the linear region shrinks as the temperature decreases.  At the lowest temperature $4.2K$, there is no visible linear regime within the measurement limit and a clear power law dependence of conductance  with  $V_{ds}$ is observed. For higher bias values, there is a tendency for the conductance to saturate which is present at all temperatures. The suppression of conductance is the result of electron backscattering  due to emission of zone-boundary or OPs \cite{Park,Yao}. We perform the noise measurements at $4.2K$ in the the power law regime for two metallic SWNTs, one with a high contact resistance (HR) and another with a low contact resistance (LR). This was done in order to understand any influence of contact resistance on the noise analysis for the tubes. The $\alpha$ values for the LR and HR SWNT are 0.9$\pm$0.02 and 3.48$\pm$0.06 respectively. The HR-SWNT has a diameter of 1.6 nm and resistance $\sim$1M$\Omega$ and the LR-SWNT has a diameter of 2.9 nm and resistance  $\sim$100K$\Omega$.

		  In  figure 2 we show the behavior of relative noise ($A_I$)  as a function of $V_{ds}$ for both the SWNTs in the power law regime and beyond at 4.2K. The relative noise power ($S_I/I^{2+\gamma_o}$) is extracted  using the generalized Hooge's relation for both SWNTs in the frequency interval  20-40 Hz. The $\gamma_o$ values for the LR and HR-SWNT are 0 and -1 respectively. From the plots it is clear that the relative noise power remains constant (shown by the yellow band) up to $\sim$ 170mV, i.e, up to the power law region for both the tubes.  Recall that the relative noise is also expected to be a constant in the power-law regime according to equation \ref{eq:pwrco}. The situation is similar to the linear regime where the relative noise ($S_I/I^2$) is a constant which is a consequence of Ohm's law.  This is the main result of our work where we show that even in this highly non-linear regime the relative noise behavior is the same as in the Ohmic regime. The analysis can be extended well above the power law regime also. Beyond the power-law region ($\sim$ 170 mV), the relative noise power seems to decrease with increasing $V_{ds}$. It is well known from various transport and optical measurements \cite{Dresselhaus,Jorio} that  backscattering by OP is responsible for saturation of conductance in metallic SWNTs at phonon energies of $\hbar\Omega$ = 0.16--0.17 V. This energy value coincides with the energy where the conductance starts to saturate and the relative noise ($A_I$) magnitude begins to reduce. This suggests that OP scattering causes the reduction in relative noise beyond the power-law regime. The detailed mechanism for the reduction in noise magnitude is explained later.  Although contacts play an important role for producing noise in CNT devices, identical qualitative behavior of relative noise for both the tubes proves that the origin of noise is indeed intrinsic. Furthermore, in the high bias regime  we can effectively neglect the contribution of contacts \cite{Back, Yao} to the overall noise magnitude.

\bigskip			 
\begin{figure}
\includegraphics[width=8cm]{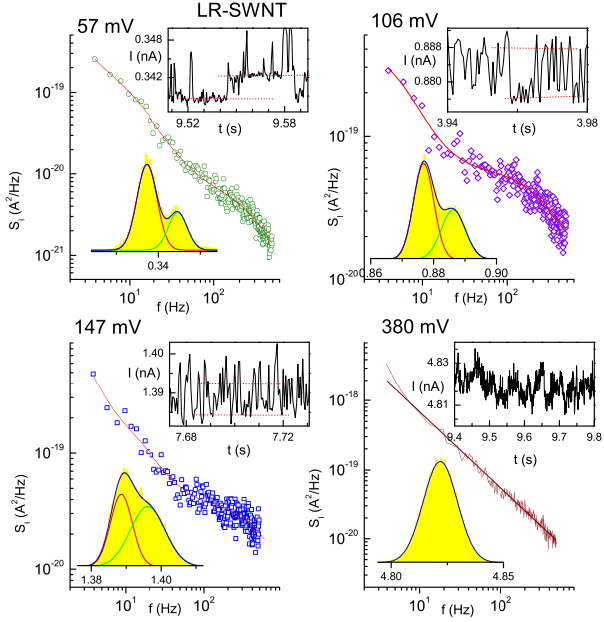}
\caption{(color online) Change in nature, distribution and PSD of noise with increase in $V_{sd}$ (a) Time series (top inset), distribution of the fluctuations (bottom inset) and the PSD of the fluctuations for a bias of (a) 57mV (b)  106 mV (c)  147 mV  and (d) 380 mV. The two levels in the time series are indicated by red dashed lines. The PSD of the fluctuations can be fitted by lorentzian for lower bias values. At high bias values   PSD is characteristic of  $1/f$ noise. }
\label{fig.3}
\end{figure}

		    In case of SWNTs two-level fluctuations (RTN) has  been observed at 4.2K in previous studies \cite{Liu} which arise due to carrier trapping and detrapping from individual defects located in the oxide. In this case  also the time series of  fluctuations for LR-SWNT show distinct switching between two voltage levels. The distinction between the two levels in the time series fades gradually with increase in bias and at the highest $V_{ds}$ it completely vanishes as shown in figure 3. The current power spectral density can be fitted by the sum of two lorentzians  up to $V_{ds}$= 210   $mV$ after which the spectra has $1/f$ character. Looking at the evolution of time series, the noise in LR-SWNT can  be thought to arise from a combination of RTN from trap fluctuations and OP scattering. However, the blurring of two distinct levels in the  time-series as well as distribution of fluctuations  anticipate the onset of OP scattering (0.17  $V$) effectively ruling out any correlation between the two noise sources. For the HR-SWNT, we do not observe any RTN and the noise remains $1/f$ throughout the measurement range. This is due to the reason that in HR-SWNTs noise rises above the background only after a relatively high bias value ($\sim$80 mV) making observation of RTN difficult which is usually observable at low bias values \cite{Malchup}.

		Recall that the unusual reduction in the magnitude of relative noise  with bias shown in figure 2 has been ascribed to OP scattering. We develop a noise model based on the Hooge's mobility fluctuation approach  to gain a more quantitative understanding of the phenomena. One can use Boltzmann transport theory \cite{Park} to calculate the current in metallic SWNTs at high bias. At high bias, the current can be approximated as $I=B l_{\rm eff}/L$, where $B$ is a constant, $L$ is the length of the CNT and  $l_{\rm eff}$ the effective mean free path.  It is well known that low frequency noise in SWNTs originate from mobility fluctuations \cite{Liu,Ishigami}. In this case, the modulation of the effective mean free path (mfp) due to OP scattering  translates into mobility fluctuations of the device. By Mathissen's rule, the effective \textit{mfp} is given by $l_{\rm eff}^{-1}$=$l_e^{-1}$+$l_{\rm hp}^{-1}$, where $l_e$ is the elastic scattering \textit{mfp} and $l_{\rm hp}$ is the \textit{mfp} of backscattering phonons. Therefore, the only possibility is that the noise  at high bias for SWNTs originates due to slow fluctuations in  $l_{hp}$, caused by scattering with OPs. This modulation of path lengths in turn translates into mobility fluctuations of the device. The fluctuations in current $I$, is then given by, $\delta I= B \delta l_{\rm eff}/L$. At a given instant if the SWNT has $N$ electrons with mean free path lengths $l_i(i=1,.....,N)$ then an effective \textit{mfp} is given by $l_{\rm eff}=(\sum l_i)/N$.  These  slow fluctuations in the path lengths $\delta l_{i}$ would naturally be related to the scattering rate of the electrons with the OPs. The relative noise power spectra $A_I$ is then given by,
\begin{equation}
S_I= B^2 \frac{(\sum S_{l_i})}{L^2 N^2},
A_I=[\frac{S_{l_{\rm hp}}}{l_{hp}^{2+\gamma_o}}] \frac{1}{N}
\end{equation}
as $(\sum S_{l_i})/N = S_{l_{\rm hp}}$ and $S_I \propto I^{2+\gamma_o}$.

			Due to the presence of a component of $l_{\rm hp}$ term in each $l_i$, the low frequency fluctuations will have a signature from OP scattering which would be proportional to the  the scattering rate of the electrons with the OPs even though OP scattering takes place in the time scale of pico-seconds. Using Fermi's golden rule and considering only emission processes, the decay time for an electron in state $\bf k$, band $l$ and energy $\epsilon_{{\bf \bf k}l}$ to another electronic band  $l'$ with energy $\epsilon_{({\bf \bf k+q})l'}$ is given by \cite{Lazzeri,Popov} :   			 
\begin{equation}
\frac{1}{\tau} = \sum_{\eta} \frac{\pi}{M\omega_{{\bf q}\eta}} |D_{({
\bf k}+{ \bf q})l',{ \bf k}l}|^2 \rho[\epsilon_{({ \bf k}+{ \bf
q})l'}]~(n_{ {\bf -q}\eta}+1).
\label{eq3}
\end{equation}
Here $\bf q$ is the phonon wavevector in branch $\eta$ with energy $\hbar \omega_{{\bf q}\eta}$, $|D|$ the electron-phonon-coupling (EPC) strength, $\rho$ is the density of states and $n$ is the phonon occupation factor given by $n_{{\bf q}\eta}=[exp(\hbar\omega_{{\bf q}\eta}/k_BT)-1]^{-1}$.  For $eV_{ds}$$\geq$$\hbar\omega_q$, where $\omega_q$ is the optical phonon frequency, a key role is played by  phonon occupation number in  limiting the high-field transport as well as low frequency noise in nanotubes. In this limit, assuming the energies are shifted such  as $E\stackrel{}{\rightarrow}eV_{ds}+E_F$ by applied bias $V_{ds}$, the scattering rate becomes \cite{Back}:  $\tau(V_{ds})^{-1}$$\propto$ $\rm exp{-[(eV_{ds}+E_F)-\hbar\omega_q]/k_bT}$. We find that the noise can be fit remarkably well by the simple exponential function of $V_{ds}$ beyond the power-law region as shown in figure 2. Physically, the low frequency noise in metallic CNTs caused by mobility fluctuations is related to the OP scattering rate through the phonon occupation number.

\begin{figure}
\includegraphics[height=5.5 cm]{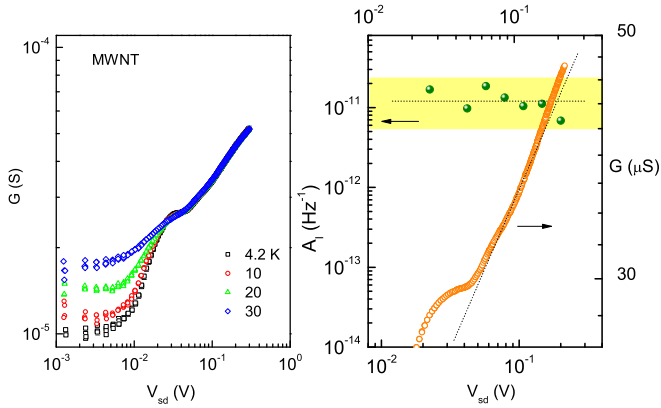}
\caption{(color online) (a) Conductance $G$ of a MWNT as a function of source-drain voltage $V_{sd}$ at different temperatures. A power law above $V_{ds}$= 10 $mV$ is evident at 4.2K. (b) Relative noise $A_I$ versus source-drain bias $V_{sd}$ at 4.2K. The bias independence of the relative noise can be observed in the power-law regime which is indicated by the yellow band.}
\label{fig.4}
\end{figure}

            Returning to the problem of noise in power law regime we  perform further noise measurements in MWNTs which have a cleaner and much larger power-law regime than SWNTs. In MWNTs the power-law regime extends up to a much higher source drain bias, $V_{ds}$ due to their ability to carry much higher  currents.  In figure 4a the power-law dependence of conductance on drain-source voltage $V_{ds}$ at different temperatures in an MWNT device is shown. Note that the power-law spans the entire measurement range (more than a decade) and there is no conduction saturation due to optical phonon scattering as the threshold for OP scattering being at a much higher bias in MWNTs compared to SWNTs\cite{Collins,Chiu,Bourlon}. The origin of the unusual kink in conductance seen in the figure is also not clear  and might be caused due to some disorder in the tube. The relative noise ($A_I$) in the power-law region for the MWNT is shown in figure 4b. In  agreement with equation (3) we observe that the relative noise ($A_I$) remains constant with voltage ($V_{ds}$) in the power-law regime.  The PSD of noise also shows a  $1/f$ character in the entire power law regime in the MWNT. Measurements in MWNTs also once again clearly show that noise in a power-law regime behaves similar to that in linear regime.

     In conclusion, we have presented a framework for studying current fluctuations in the non-Ohmic regime. This novel study was demonstrated by measuring current noise in the power-law region and beyond in SWCNT and MWCNTs. It is shown that the generalized relative noise in this highly non-linear regime remains independent of bias similar to the linear regime. As a utility of this study we show that by measuring noise beyond the power-law regime that mobility fluctuations of charge carriers in the tube caused by scattering with OPs is dominant source of noise in CNTs in this regime.  We propose a mobility fluctuation model to relate mobility fluctuations to the phonon occupation number. Our experimental results have widespread ramifications on noise characterization for high bias interconnect applications of nanoelectronic devices which operate mostly in the non-linear regime.

\textbf{Acknowledgments} The authors would like to thank C. G. Rocha and  K.K. Bardhan for useful discussions and comments. D.T. would like to acknowledge CIMO for a grant. O.H. acknowledges Jenny and Antti Wihuri Foundation and Finnish Cultural Foundation for supporting this work.


\begin{thebibliography} {deeps Label}


\bibitem{Kogan}  Sh. Kogan, \text{Electronic Noise and Fluctuations in Solids},  (Cambridge University Press, Cambridge, UK, 1996).

\bibitem{Rammal} R. Rammal, and A.M. Tremblay, \text{ Phys. Rev. Lett.} \textbf{58}, 415-418 (1987). 


\bibitem{Zhou} Z. Zhou, K. Xiao, R. Jin, D. Mandrus, J. Tao, D. B. Geohegan, and S. Pennycook,  \text{ App. Phys. lett.} \textbf{90}, 193115 (2007).     

\bibitem{Liu1} F. Liu , M. Bao, K.L. Wang, C. Li, B. Lei, and C. Zhou,  \text{ App. Phys. lett.} \textbf{86}, 213101 (2005).    

\bibitem{Venkataraman} L. Venkataraman,  Yeon Suk H. Hong, and P. Kim ,\text{ Phys. Rev. Lett.} , \textbf{96} , 076601, (2006). 

   
\bibitem{Bockrath} M. Bockrath, D. H. Cobden, J. Lu, A. G. Rinzler, R. E. Smalley, L. Balents, and P. L. McEuen,  \text{Nature}, \textbf{397}, 598-601 (1999).   
 
\bibitem{Kane} C. Kane, L. Balents, and M. P. Fisher, \text{ Phys. Rev. Lett.} \textbf{79}, 5086-5089 (1997).

\bibitem{Tarkiainen} R. Tarkiainen, M. Ahlskog, J. Penttilä, L. Roschier , P. Hakonen, M. Paalanen, and E. Sonin, \text{ Phys. Rev. B} \textbf{64},195412  (2001).
 

\bibitem{Weissman} M. B. Weissman, \text{Rev. Mod. Phy.}, \textbf{60}, 537-571 (1988).    

\bibitem{Talukdarrsi} D. Talukdar, R.K. Chakraborty, S. Bose, and K. K. Bardhan, \text{Rev. Sci. Instr.}, \textbf{82}, 013906 (2011).

\bibitem{BardhanUN} U. N. Nandi, C. D. Mukherjee, and K. K. Bardhan, \text{ Phys. Rev. B} \textbf{54},12903 (1996).

\bibitem{Hooge} F. N. Hooge, \text{Phys. Lett. A}, \textbf{29}, 139 (1969).


\bibitem{Bardhan} K. K. Bardhan, C. D. Mukherjee, and U. N. Nandi,   \text{AIP Conf. Proc.} \textbf{800}, 109-117 (2005).   


\bibitem{Park}  J. Y. Park, S. Rosenblatt, Y. Yaish, V. Sazonova, H. Üstünel, Braig S., and McEuen P. L., \text{ Nano Letters} \textbf{4}, 517-520 (2004).


 
\bibitem{Yao} Z. Yao, C. L. Kane, and C. Dekker,\text{ Phys. Rev. Lett.} \textbf{84}, 2941 (2000). 
       

\bibitem{Back} J. H. Back, C. L. Tsai, S. Kim, S. Mohammadi, and M. Shim,\text{ Phys. Rev. Lett.} \textbf{103}, 215501 (2009). 

\bibitem{Dresselhaus} M. S. Dresselhaus, G. Dresselhaus, R. Saito, A. Jorio, \text{Phys. Rep.} \textbf{409}, 47 (2005). 

\bibitem{Jorio} A. Jorio,  M. A. Pimenta, A. G. Souza Filho, R. Saito, G. Dresselhaus and M. S. Dresselhaus,\text{New J. Phys.}  \textbf{5}, 139 (2003). 
 
  
\bibitem{Liu} F. Liu, K. L. Wang, D. Zhang, and C. Zhou, \text{ App. Phys. lett.} \textbf{89}, 063116-063116 (2006).  
       
\bibitem{Malchup} S. Malchup, \text{ J App. Phys.} \textbf{25}, 341 (1954).  

\bibitem{Ishigami} M. Ishigami,  J. H. Chen,  E. D. Williams, D. Tobias, Y. F. Chen, and M. S. Fuhrer, \text{ App. Phys. lett.} \textbf{88}, 203116-203116 (2006). 


\bibitem{Lazzeri}  M. Lazzeri, S. Piscanec, F. Mauri, A.C. Ferrari, and J. Robertson,\text{ Phys. Rev. Lett.} \textbf{95}, 236802 (2005).   
	
\bibitem{Popov} V. N. Popov, and P. Lambin, \text{ Phys. Rev. B} \textbf{74}, 075415 (2006).  


\bibitem{Collins} P. G. Collins, M. Hersam, M. Arnold, R. Martel, and P. Avouris,\text{ Phys. Rev. Lett.} \textbf{86}, 3128-3131 (2001).  

\bibitem{Chiu}  H. Y. Chiu, V. V. Deshpande, H. C. Postma,  C. N. Lau, C. Miko ,  L. Forro, and  M. Bockrath,\text{ Phys. Rev. Lett.} \textbf{95}, 226101 (2005).   

\bibitem{Bourlon}  B. Bourlon,  D. C. Glattli, B. Plaçais ,  J. M. Berroir,  C. Miko,  L. Forro, and A. Bachtold,\text{ Phys. Rev. Lett.} \textbf{92}, 26804 (2004).     

 
 
\end{thebibliography}
\end{document}